
\input harvmac
\input epsf
\catcode`\:=11
\newcount\bm:counta \newcount\bm:countb 
\newcount\bm:countc \newcount\bm:countd
\newtoks\bm:tok
\newif\ifbm:delim
\def\thehex#1{\ifcase\the#1 0\or 1\or 2\or 3\or 4\or 5\or 6\or 7\or
8\or 9\or A \or B\or C\or D\or E\or F\fi}%
\def\test#1#2{\ifcat#1#2\message{True}\else\message{False}\fi}
\newtoks \bm:savedtoks  \bm:savedtoks{}%
\def\bm:empty{\relax}%
\let\bm:save=\bm:empty
\newif\ifbm:cdr
\def\bm:split#1#2\bm:empty{%
    \def\bm:car{#1}\def\bm:cdr{#2\relax}%
    \expandafter\ifx\expandafter\relax\bm:cdr\bm:cdrfalse
    \else\bm:cdrtrue
    \fi}%
\newif\ifbm:found
\def\bm:in#1\find:#2\this:{%
    \def\find:##1#2##2##3\find:{%
        \ifx\bm:in##2\bm:foundfalse
        \else\bm:foundtrue
        \fi}%
    \find:#1#2\bm:in\find:}%
\let\when=\relax \let\use=\relax
%
%
\newif\ifboldwarning
\def\bm:message#1{{\newlinechar=`^^J
\immediate\write16{\string\bold\space warning on line
                                             \the\inputlineno^^J#1^^J}}}%
%
\def\latex:adjust{\expandafter\ifx\the\textfont0\csname twlrm\endcsname
                                           \def\bm:scale{1200}%
                              \else\expandafter\ifx
                                 \the\textfont0\csname elvrm\endcsname
                                                 \def\bm:scale{1095}%
                                               \else\def\bm:scale{1000}%
                                               \fi
                              \fi}%
\latex:adjust
\newdimen\bm:sevensize \newdimen\bm:fivesize
\bm:sevensize=.007pt \bm:fivesize=.005pt
\bm:sevensize=\bm:scale\bm:sevensize
\bm:fivesize=\bm:scale\bm:fivesize
\font\tenbf=cmbx10 scaled \bm:scale
\font\sevenbf=cmbx7 at \the\bm:sevensize
\font\fivebf=cmbx5 at \the\bm:fivesize
\textfont\bffam=\tenbf
\scriptfont\bffam=\sevenbf
\scriptscriptfont\bffam=\fivebf
:bit=cmbxti10 scaled \bm:scale
:bit=cmbxti10 at \the\bm:sevensize
:bit=cmbxti10 at \the\bm:fivesize
:scale
 at \the\bm:sevensize
 at \the\bm:fivesize
:bsf=cmssbx10 scaled \bm:scale
:bsf=cmssbx10 at \the\bm:sevensize
:bsf=cmssbx10 at \the\bm:fivesize
:bsl=cmbxsl10 scaled \bm:scale
:bsl=cmbxsl10 at \the\bm:sevensize
:bsl=cmbxsl10 at \the\bm:fivesize
:bmit=cmmib10 scaled \bm:scale
:bmit=cmmib7 at \the\bm:sevensize
:bmit=cmmib5 at \the\bm:fivesize
\newfam\bm:bmitfam
\textfont\bm:bmitfam=\tenbm:bmit
\scriptfont\bm:bmitfam=\sevenbm:bmit
\scriptscriptfont\bm:bmitfam=\fivebm:bmit
:bsy=cmbsy10 scaled \bm:scale
:bsy=cmbsy7 at \the\bm:sevensize
:bsy=cmbsy5 at \the\bm:fivesize
\newfam\bm:bsyfam
\textfont\bm:bsyfam=\tenbm:bsy
\scriptfont\bm:bsyfam=\sevenbm:bsy
\scriptscriptfont\bm:bsyfam=\fivebm:bsy
\newtoks\alphatok \alphatok{?}%
\expandafter\def\expandafter\new:fam\expandafter{\newfam}%
\def\set:fam#1{%
    \expandafter\ifx\csname #1fam\endcsname\relax
                    \expandafter\new:fam\csname #1fam\endcsname
                    \edef\alphafam{\csname #1fam\endcsname}%
                    \edef\set:fonts{%
                         \global\textfont\alphafam=\csname ten#1\endcsname
                         \global\scriptfont\alphafam=\csname
                                                seven#1\endcsname 
                         \global\scriptscriptfont\alphafam=\csname
                                                        five#1\endcsname}%
                    \set:fonts
                \fi
}%
\def\declare:alpha#1{%
    \set:fam{#1}%
    \expandafter\edef\csname math#1\endcsname{{%
                     \noexpand\if?\noexpand\the\noexpand\alphatok
                                  \global\noexpand\bm:savedtoks
                            {\noexpand\csname math:#1\noexpand\endcsname}%
                                  \noexpand\aftergroup\noexpand\bm:getarg
                     \noexpand\else\errmessage{You're already inside a
                                   \noexpand\expandafter\noexpand\string
                                   \noexpand\the\alphatok{..} - don't
                                   even think about it!}%
                     \noexpand\fi}}%
    \expandafter\def\csname math:#1\endcsname##1{{%
                     \alphatok\expandafter{\csname math#1\endcsname}%
                     \fam\csname #1fam\endcsname
                     \let\boldletter=\relax ##1}}}%
%
\chardef\rmfam=0
\chardef\mitfam=1
\chardef\calfam=2
\declare:alpha{rm}%
\declare:alpha{it}%
\declare:alpha{sl}%
\declare:alpha{tt}%
\declare:alpha{bf}%
\declare:alpha{mit}%
\declare:alpha{cal}%
\declare:alpha{sf}
\declare:alpha{bm:bmit}
\declare:alpha{bm:bsy}
\def\default{\fam=-1 \def\boldletter##1{{\bf ##1}}}%
\let\mathbm=\mathbm:bmit :bsy
\let\mathbm:bcal=\mathbm:bsy
\def\bold#1{{
    \let\bm:currentsymbol=\relax
    \bm:split#1\bm:empty
    \ifbm:cdr\toks0{}\loop\toks0\expandafter\expandafter\expandafter{\expandafter\the\expandafter\toks0\expandafter\noexpand\expandafter\bold\expandafter{\bm:car}}\expandafter\bm:split\bm:cdr\bm:empty
                      \ifbm:cdr
                      \repeat
              \toks2\expandafter{\bm:car}%
              \xdef\bm:out{\the\toks0
                           \noexpand\bold\expandafter{\the\toks2}}%
              \aftergroup\bm:out 
    \else\ifx\bold#1\aftergroup\bold
         \else\ifmmode\bm:select{#1}%
                      \ifx\bm:save\bm:empty
                          \xdef\bm:out{\global\bm:savedtoks{}%
                             \the\bm:savedtoks\noexpand\bm:currentsymbol}%
                          \aftergroup\bm:out
                      \else\global\bm:savedtoks\expandafter\expandafter
          \expandafter{\expandafter\the\expandafter\bm:savedtoks\bm:save}%
                            \ifx\bm:currentsymbol\relax\aftergroup\bold
                            \else\let\test=F%
                                 \edef\argtest{\noexpand\bm:in
                                      \meaning\bm:currentsymbol
                                      \noexpand\find:
                                      \string\mathaccent
                                      \noexpand\this:}%
                                 \argtest
                                 \ifbm:found \let\test=T%
                                 \fi
                                 \edef\argtest{\noexpand\bm:in
                                      \meaning\bm:currentsymbol
                                      \noexpand\find:
                                      \string\radical
                                      \noexpand\this:}%
                                 \argtest
                                 \ifbm:found \let\test=T%
                                 \fi
                                 \if T\test\aftergroup\bm:getarg
                                 \else\aftergroup\bm:dumpchars
                                 \fi
                            \fi
                      \fi
              \else\errmessage{\string\bold\space should be used in
                               math mode only}%
              \fi
          \fi
    \fi
    }}%
\def\bm:getarg#1{{%
    \ifx#1\bold\aftergroup\bold
    \else\toks0{#1}\xdef\bm:out{\global\bm:savedtoks{}%
                        \the\bm:savedtoks{\the\toks0}}%
         \aftergroup\bm:out
    \fi}}%
\def\bm:dumpchars{{%
    \xdef\bm:out{\global\bm:savedtoks{}\the\bm:savedtoks}%
    \aftergroup\bm:out}}%
\def\bm:select#1{%
    \expandafter\bm:in\boldspecials\find:#1\this:
    \ifbm:found \def\when##1\use##2;{%
                    \ifx##1#1\xdef\bm:currentsymbol{\noexpand##2}%
                    \else\ifx##2#1\xdef\bm:currentsymbol{\noexpand##2}\fi
                    \fi}%
                \boldspecials
    \else\if\noexpand#1\relax                    
            \let\test=F%
            \edef\chartest{\noexpand\bm:in\meaning#1\noexpand\find:
                                         \string\mathchar\noexpand\this:}%
            \chartest
            \ifbm:found\let\test=T%
            \fi
            \edef\chartest{\noexpand\bm:in\meaning#1\noexpand\find:
                                           \string\char\noexpand\this:}%
            \chartest
            \ifbm:found \let\test=T%
            \fi                                  
            \if T\test 
          \expandafter\ifx\csname bold\string#1\endcsname\relax 
                          \edef\bm:process{\noexpand\defboldsymbol
                                           {\noexpand#1}%
                                           {\noexpand\mathchar}%
                                           {\the#1}{-1}}%
                          \bm:process
                       \fi \global\expandafter\let\expandafter
                           \bm:currentsymbol\expandafter=%
                           \csname bold\string#1\endcsname
            \else 
          \expandafter\ifx\csname bold\string#1\endcsname\relax 
                          \ifboldwarning\bm:message{Skipping \string#1.}%
                          \fi
                                        \def\bm:save{#1}%
                      \else\toks4\expandafter{%
                           \csname bold\string#1\endcsname}%
                           \edef\bm:save{\the\toks4}%
                           \global\expandafter\let\expandafter
                           \bm:currentsymbol\expandafter=%
                           \the\toks4
                      \fi
            \fi
         \else                                   
              \ifcat\noexpand#1A\gdef\bm:currentsymbol{\boldletter{#1}}%
              \else\ifcat\noexpand#1>            
             \expandafter\ifx\csname bold\string#1\endcsname\relax
                             \edef\bm:process{\noexpand\defboldsymbol
                                              {\noexpand#1}%
                                              {\ifnum\the\delcode`#1>-1
                                                 \noexpand\delimiter
                                               \else\noexpand\mathchar
                                               \fi}%
                                              {\the\mathcode`#1}%
                                              {\the\delcode`#1}}%
                             \bm:process         
                          \fi\global\expandafter\let
                             \expandafter\bm:currentsymbol
                             \expandafter=\csname 
                             bold\string#1\endcsname
                   \else\ifboldwarning\bm:message{Skipping \string#1.}%
                        \fi
                                      \def\bm:save{#1}%
                   \fi
              \fi
         \fi
    \fi}%
\def\defboldsymbol#1#2#3#4{%
    \bm:tok={#1}%
 \expandafter\ifx\csname bold\string#1\endcsname\relax
             \else\ifboldwarning\bm:message{Redefining
                                                \string\bold\string#1.}%
                  \fi
             \fi
    \bm:counta=#3
    \bm:countd=#4
    \ifnum\the\bm:counta="8000
           \expandafter\xdef\csname bold\string#1\endcsname{#1}%
    \else\ifx#2\delimiter \bm:delimtrue
         \else\ifx#2\radical \bm:delimtrue
              \else \bm:delimfalse
              \fi
         \fi
         \bm:countc=\bm:counta
         \divide\bm:countc by "1000             
         \advance\bm:counta by -\expandafter"\thehex\bm:countc 000
         \ifbm:delim\ifnum\the\bm:countd>-1     
                          \begingroup \bm:counta=\bm:countd
                          \divide\bm:counta by "1000
                              \begingroup
                              \multiply\bm:counta by "1000
                              \global\advance\bm:countd by
                                                      -\the\bm:counta
                              \endgroup
                          \bold:mathrecode
                          \multiply\bm:counta by "1000
                              \begingroup
                              \bm:counta=\bm:countd
                              \bold:mathrecode
                              \global\bm:countd=\bm:counta
                              \endgroup
                          \global\advance\bm:countd by \the\bm:counta
                          \endgroup
                      \else\ifboldwarning\bm:message{\the\bm:tok\space
                        is not a \string#2.^^JDoing the obvious thing..}%
                           \fi
                      \bold:mathrecode
                      \bm:countd=\bm:counta
                      \multiply\bm:counta by "1000
                      \advance\bm:countd by \the\bm:counta
                      \fi
                      \advance\bm:countd by \expandafter"\thehex
                                                        \bm:countc000000
                      \expandafter\xdef\csname bold\string#1\endcsname
                               {#2\the\bm:countd}%
          \else\bold:mathrecode
               \advance\bm:counta by \expandafter"\thehex\bm:countc 000
               \ifx#2\mathchar
                   \global\expandafter\mathchardef\csname
                                bold\string#1\endcsname=\the\bm:counta
               \else\expandafter\xdef\csname bold\string#1\endcsname
                               {#2\the\bm:counta}%
               \fi
          \fi
    \fi
}%
\def\bold:mathrecode{
      \bm:countb=\bm:counta
      \divide\bm:countb by "100
      \advance\bm:counta by -\expandafter"\thehex\bm:countb 00
      \ifcase\the\bm:countb
              \bm:countb=\the\bffam
      \or     \bm:countb=\the\bm:bmitfam
      \or     \bm:countb=\the\bm:bsyfam
      \else   \ifboldwarning\ifbm:delim\bm:message{Lack of bold
                                        extension fonts means
                                        \string\bold\the\bm:tok\space may
                                        not be bold.}%
                             \else\bm:message{Sorry, there just aren't the
                                   fonts for \string\bold\the\bm:tok.}%
                             \fi
              \fi 
      \fi
\advance\bm:counta by \expandafter"\thehex\bm:countb 00
                    }%
\def\DeclareBoldMacro#1#2#3{
    \bm:counta=#3 \bm:countd=\bm:counta
    \ifnum\the\bm:counta>"7FFF 
          \divide\bm:counta by "1000
    \else
          \multiply\bm:countd by "1000
    \fi
    \bm:countc=\bm:counta
    \divide\bm:countc by "1000
    \advance\bm:countd by -"\thehex\bm:countc 000000
    \edef\bm:process{\noexpand\defboldsymbol{\noexpand#1}{\noexpand#2}%
                     {\the\bm:counta}{\the\bm:countd}}%
    \bm:process}%
%
%
\DeclareBoldMacro{\{}{\delimiter}{"4266308}
\DeclareBoldMacro{\}}{\delimiter}{"5267309}
\DeclareBoldMacro{\langle}{\delimiter}{"426830A}
\DeclareBoldMacro{\rangle}{\delimiter}{"526930B}
\edef\FixLessThanGreaterThan{\noexpand\DeclareBoldMacro{<}%
{\noexpand\mathchar}{\the\mathcode`\<}%
\noexpand\DeclareBoldMacro{>}{\noexpand\mathchar}{\the\mathcode`\>}}%
\FixLessThanGreaterThan
\DeclareBoldMacro{\sqrt}{\radical}{"270370}
\default
\def\boldspecials{\when\mathmit\use\mathbm:bmit;\when\mathcal\use\mathbm:bcal;\when\mathrm\use\mathbf;\when\mathsf\use\mathbm:bsf;\when\mathit\use\mathbm:bit;\when\mathtt\use\mathtt;\when\mathsl\use\mathbm:bsl;\when\default\use\default;}%
\boldwarningtrue
\catcode`\:=12

\magnification\magstep1
\parskip 6pt
\newdimen\itemindent \itemindent=32pt
\def\textindent#1{\parindent=\itemindent\let\par=\resetpar%
\indent\llap{#1\enspace}\ignorespaces}

\let\oldpar=\par
\def\resetpar{\oldpar\parindent=20pt\let\par=\oldpar}

\font\ninerm=cmr9 \font\ninesy=cmsy9
\font\eightrm=cmr8 \font\sixrm=cmr6
\font\eighti=cmmi8 \font\sixi=cmmi6
\font\eightsy=cmsy8 \font\sixsy=cmsy6
\font\eightbf=cmbx8 \font\sixbf=cmbx6
\font\eightit=cmti8
\def\eightpoint{\def\rm{\fam0\eightrm}
  \textfont0=\eightrm \scriptfont0=\sixrm \scriptscriptfont0=\fiverm
  \textfont1=\eighti  \scriptfont1=\sixi  \scriptscriptfont1=\fivei
  \textfont2=\eightsy \scriptfont2=\sixsy \scriptscriptfont2=\fivesy
  \textfont3=\tenex   \scriptfont3=\tenex \scriptscriptfont3=\tenex
  \textfont\itfam=\eightit  \def\it{\fam\itfam\eightit}%
  \textfont\bffam=\eightbf  \scriptfont\bffam=\sixbf
  \scriptscriptfont\bffam=\fivebf  \def\bf{\fam\bffam\eightbf}%
  \normalbaselineskip=9pt
  \setbox\strutbox=\hbox{\vrule height7pt depth2pt width0pt}%
  \let\big=\eightbig  \normalbaselines\rm}
\catcode`@=11 %
\def\eightbig#1{{\hbox{$\textfont0=\ninerm\textfont2=\ninesy
  \left#1\vbox to6.5pt{}\right.\n@space$}}}
\def\vfootnote#1{\insert\footins\bgroup\eightpoint
  \interlinepenalty=\interfootnotelinepenalty
  \splittopskip=\ht\strutbox %
  \splitmaxdepth=\dp\strutbox %
  \leftskip=0pt \rightskip=0pt \spaceskip=0pt \xspaceskip=0pt
  \textindent{#1}\footstrut\futurelet\next\fo@t}
\catcode`@=12 %
\def \l{\langle}
\def \r{\rangle}
\def \de{\delta}
\def \si{\sigma}

\def \pr{\partial}

\def \bx{{\bf x}}
\def \bb{{\bf b}}
\def \he{{\hat e}}
\def \hx{{\hat x}}
\def \hy{{\hat y}}
\def \hz{{\hat z}}
\def \hn{{\hat n}}
\def \hv{{\hat v}}
\def \d{{\rm d}}

\def \half{{\textstyle {1 \over 2}}}

\def \hh{{1\over 2}}
\def \ts{ \textstyle}

\def \A{{\cal A}}
\def \B{{\cal B}}

\def \O{{\cal O}}
\def \M{{\cal M}}
\def \V{{\cal V}}
\def \R{{\cal R}}

\def \bb{{\bf b}}
\def \bn{{\bf n}}
\def \bv{{\bf v}}
\def \bw{{\bf w}}
\def \ep{\epsilon}

\font \bigbf=cmbx10 scaled \magstep1

\lref\hughtwo{J. Erdmenger and H. Osborn, Nucl. Phys. {\bf B483} (1997)
431; hep-th/9605009.}
\lref\hughone{H. Osborn and A. Petkou,
    Ann. Phys. {\bf 231} (1994) 311; hep-th/9307010.}
\lref\jil{A. Cappelli, D. Friedan  and J.I. Latorre,
    Nucl. Phys. {\bf B352} (1991) 616\semi
 A. Cappelli, J.I. Latorre and X. Vilasis-Cardona,
    Nucl. Phys. {\bf B376} (1992) 510; hep-th/9109041.}
\lref\Visser{M. Visser, Phys. Rev. {\bf D54} (1996) 5103, 5116, 5123;
gr-qc/9604007,9604008,\hfil\break9604009 and gr-qc/9703001.}
\lref\fulling{S.M. Christensen and S.A. Fulling, Phys. Rev. {\bf D15} (1977)
2088}
\lref\Glaser{H. Epstein, V. Glaser and A. Jaffe, Nuovo Cimento {\bf 36} (1966)
1016.}
\lref\cardy{J.L. Cardy, Phys. Lett. {\bf B215} (1988) 749.}
\lref\Ford{L.H. Ford and T.A. Roman, Phys. Rev.  {\bf D46} (1992) 1328,
{\bf D51} (1995) 4277,
{\bf D53} (1996) 1988, {\bf D55} (1997) 2082; gr-qc/9607003\semi
M.J. Pfenning and L.H. Ford, Phys. Rev.  {\bf D55} (1997) 4813;
gr-qc/9608005\semi
{\'E}.{\'E}. Flanaghan, gr-qc/9706006.}
{\nopagenumbers
\rightline{DAMTP/97-1}
\rightline{hep-th/9703196}
\vskip 2truecm
\centerline {\bigbf Modified Weak Energy Condition for the Energy Momentum} 
\vskip 5pt
\centerline {\bigbf Tensor in Quantum Field Theory}
\vskip 2.0 true cm
\centerline {Jos\'e I. Latorre* and Hugh Osborn**\footnote{}{emails:
{\tt latorre@ecm.ub.es} and {\tt ho@damtp.cam.ac.uk}}}
\vskip 12pt
\centerline {*\ Departament d'Estructura i Constituents de la Mat\`eria,}
\centerline {Diagonal 647, 08028 Barcelona, Spain}
\vskip 8pt
\centerline {**\ Department of Applied Mathematics and Theoretical Physics,}
\centerline {Silver Street, Cambridge, CB3 9EW, England}
\vskip 2.0 true cm

{\eightpoint
\parindent 1.5cm{

{\narrower\smallskip\parindent 0pt
The weak energy condition is known to fail in general when applied 
to expectation values of the the energy momentum tensor
in flat space quantum field theory.
It is shown how the usual counter arguments against its
validity are no longer applicable if the states $|\psi \r$ for which
the expectation value is considered are
restricted to a suitably defined subspace.
A possible natural restriction on $|\psi \r$ is
suggested and illustrated by two quantum mechanical examples
based on a simple perturbed harmonic oscillator Hamiltonian.
The proposed alternative quantum weak energy condition
is applied to states formed by the action of scalar, vector and the energy
momentum tensor operators on the vacuum. We assume conformal invariance in
order to determine almost uniquely three-point functions
involving the energy momentum tensor in terms of a few parameters. The
positivity conditions lead to non trivial inequalities for these parameters.
They are satisfied in free field theories, except in one case
for dimensions close to two. Further restrictions on $|\psi \r$
are suggested which remove this problem. The inequalities which
follow from considering the state formed by applying the energy momentum
tensor to the vacuum are shown to imply that the coefficient of the
topological term in the expectation value of the trace of the energy momentum
tensor in an arbitrary curved space background is positive, in accord with
calculations in free field theories.

\narrower}}
\vfill
\eject}}
\pageno=1

\newsec{Introduction}

In classical general relativity various positivity conditions on the
energy momentum tensor for matter play an essential role in the proof of
singularity, and other, theorems (see
\ref\hawking{S.W. Hawking and G.F.R. Ellis, The Large Scale Structure of
Space-time, Cambridge University Press, Cambridge 1973, chap. 4.}). 
The simplest
and perhaps most natural such inequality is the weak energy condition which
asserts that
$n^\mu n^\nu T_{\mu\nu}(x) \ge 0$ for $n^\mu$ any timelike vector at $x$. In
a quantum theory the natural extension of such a condition is to require
that it should be true for the expectation value of
$n^\mu n^\nu T_{\mu\nu}(x)$ where $T_{\mu\nu}$ is the operator representing
the energy momentum tensor which is defined in any quantum field theory.
However it was soon realised \Glaser\ that such
a condition must in general fail  even on topologically trivial space-times
(see the appendix in
\ref\davies{P.C.W. Davies and S.A. Fulling, Proc. Roy. Soc. {\bf A356}
(1977) 237.}), apart from the possibility of negative energy densities
in the Casimir effect or in the neighbourhood of the event horizon of
black holes \Visser.

A version of the argument \davies\ showing this follows simply in the context of
elementary quantum mechanics by considering an hermitian operator $T$
and a state $| 0 \rangle$ such that
\eqn\ione{
\l 0 | T | 0 \rangle = 0 \, , \qquad  T | 0 \rangle \ne 0 \, .
}
It is then evident that the positivity condition,
\eqn\itwo{
\l \psi | T | \psi \r \ge 0 \, ,
}
cannot be true for all states $| \psi \r$. This is perhaps obvious by virtue
of the standard variational principle for determining the lowest eigenvalue
of $T$ but a formal proof may be obtained by considering
\eqn\ithree{
| \psi \r = | 0 \rangle + \ep | \phi \r \, .
}
Then, assuming $\l 0 | T | \phi \r \ne 0$,
\eqn\ifour{
\l \psi | T | \psi \r = 2\ep \, {\hbox{\it Re}} \l 0 | T | \phi \r +
\ep^2 \, \l \phi | T | \phi \r < 0 \, ,
}
for some region of small $\ep>0$ or $\ep<0$.\footnote{${}^1$}{It is perhaps
illuminating to consider the manifestly positive operator $x^2$ in simple
quantum mechanics. If $\l \psi_\lambda | x^2 | \psi_\lambda \r = \lambda>0$,
for a normalised state $| \psi_\lambda \r$, then the above argument for
$| 0 \rangle \to | \psi_\lambda \r$ and $T\to x^2 - \lambda$ shows that
there exists a state $ | \psi_{\lambda'} \r $ with $\lambda'<\lambda$
but there is no state in the Hilbert space giving $\lambda=0$.}
However this counterexample to
the general applicability of \itwo\ fails if we impose the restriction\footnote
{${}^2$}{It would be nice if $\l \psi | T | \psi \r$ could be bounded below
in terms of $\l 0 | T | \psi \r$, on the basis of \ifour\ we considered
$\l \psi | T | \psi \r \ge - 2 |{\hbox{\it Re}}\l 0 | T | \psi \r |\, ||\psi||$
but this can also be shown to be incompatible with general principles.}
\eqn\ifive{
\l 0 | T | \psi \r = 0 \, .
}

In order to say something in a semi-classical context it is clear that
some restrictions on the applicability of classical energy conditions
are necessary. Ford and coworkers \Ford\ have suggested a non local condition
involving an integral over a timelike or possibly null geodesic and
which bounds the extent to which the expectation can be negative in terms
of the width of the time averaging.
The resulting inequalities have been verified for free field
theories on flat space-time and may be compatible with extending the proof of 
some singularity theorems to semi-classical general relativity. However
these inequalities are formulated entirely in a Minkowski space framework
and it is not all clear how they might be applied to the Wick rotated
quantum field theory on Euclidean space (any non perturbative definition
of quantum field theory tends to be initially in a Euclidean framework).

Alternatively we here postulate a non standard
quantum version of the weak energy condition which is motivated
by  considering how to make the above proof of the breakdown of the
simple extension to quantum theory inapplicable. Thus we propose applying
the condition only to a subspace of the full Hilbert space, thus
\eqn\isix{
\l \psi | n^\mu n^\nu T_{\mu\nu}(0) |\psi \r \ge 0 \, , \qquad
n^\mu = (1,\bn) \, , \quad \bn^2 \le 1 \, ,
}
for all states $|\psi \r$ satisfying
\eqn\iseven{
\l 0 | n^\mu n^\nu T_{\mu\nu}(0) |\psi \r = 0 \, ,
}
and for our purposes $|\psi \r$ is given by the action of finitely many
local field operators acting on the vacuum state $|0\r$. By considering simple
examples in appendix A we show that
there is no inherent contradiction in making this assumption with the basic
principles of quantum mechanics. In this paper we demonstrate that this 
quantum energy condition can lead to non trivial constraints
by considering its application to a quantum field theory at a conformal
invariant fixed point. The virtue of assuming conformal invariance is that
the two-point functions of quasi-primary fields are then uniquely determined
up to an overall constant and there are in general a finite set of linearly
independent expressions for the three-point function (in some cases there
is only one possible form, such as when two fields are scalars). This
analysis also allows the consideration of free massless scalar and fermion
field theories in arbitrary dimension $d$ and free vector fields for $d=4$.

In the following sections we consider in turn the restrictions imposed by the 
positivity condition \isix, subject to \iseven, for states formed by the 
action of
scalar, vector current and energy-momentum tensor operators on the vacuum.
The norms of these states are expressed in terms of the Euclidean two-point 
functions for these operators and the
matrix elements involving the energy momentum tensor may also be found 
in terms of the corresponding three-point functions containing
the energy momentum tensor as well. These Euclidean correlation functions
have been found explicitly previously up to a small number of parameters
by making use of conformal invariance \refs{\hughone,\hughtwo}. 
The lengthy algebra involved in some of the
three-point function computations is reduced significantly
by restricting the state so that the three points in the correlation function 
lie on a straight line. The positivity inequalities obtained in this fashion
are checked against the results of direct calculations for free scalar 
and fermion field theories,
which realise conformal invariance for general dimension $d$, and also for
vector field theories for $d=4$.
Applications of the general results constraining the three-point
function for the energy momentum tensor are discussed further
in the conclusion. We show how our positivity conditions derived for
flat space extend to constrain the 
coefficients of the two independent terms appearing in the trace
anomaly for the expectation value of the energy momentum tensor for
a conformal field theory in four dimensions on a general curved space
background. One has been known to satisfy a positivity condition since it
determines the overall scale of the energy momentum tensor two-point
function, while the other, which is the coefficient of the topological term in 
the trace, is known to be positive for free field theory but previously there
has been no general argument requiring this.

We should stress that our alternative weak energy condition is motivated
by seeing what might be feasible in terms of not being manifestly incorrect
in quantum field theory. Even if it is valid it is not at all clear how
it might be applied to semi-classical general relativity, although as
remarked above it does have implications concerning the energy momentum
tensor trace on curved space. Our detailed discussion is restricted to
conformally invariant quantum field theories, which includes the case of
free massless fields. While we have not undertaken detailed calculations
we believe thsat there should be no intrinsic difficulty in extending the
analysis to massive free fields.

As mentioned above we illustrate in appendix A the basic
philosophy by applying the inequalities to two simple 
quantum mechanics problems based on a perturbed harmonic oscillator.
The positivity condition is verified if the perturbation is not too
large and we are thus able to demonstrate that there is no 
inherent incompatibility
between our version of the weak energy condition and the general formalism of 
quantum mechanics. Appendix B  is devoted to proving that no further
information is obtained when our calculations are carried out
for more general states than those which lead to a collinear
coordinate configurations. This makes essential use of conformal invariance
which allows any three points to be transformed to lie on a line.

\newsec{Correlators involving scalar operators}

Let us start by recalling some basic results
on the constraints imposed by unitarity on two-point correlation
functions involving scalar operators, which are equivalent
to those following from    reflection
positivity in Euclidean space.

Initially we consider states formed by the action of a scalar field $\O$,
of dimension $\eta$, on the vacuum and take
\eqn\inine{
|\psi_\O \r = e^{-H\tau} \O(0,\bx) |0\r \, , \quad \tau >0 \, ,
}
where $H$ is the Hamiltonian given as usual in terms of the energy momentum
tensor by
\eqn\iten{
H = \int \! \d^{d-1} x \,\, T_{00}(0,\bx) \, .
}
In the conformal limit when $g^{\mu\nu}T_{\mu\nu}=0$ it is evident that
automatically $\l0| T_{\mu\nu} |\psi_\O \r =0$.
On continuation $x^0 \to -i\tau$ to Euclidean space the two-point function
for the scalar field $\O$ is given in the conformal limit by the
correlation function
\eqn\ieleven{
\l \O^E(x) \, \O^E(0) \r = C_\O \, {1\over x^{2\eta}} \, ,
}
where $x^2$ is defined by using  the standard Euclidean metric.
The norm of the state $|\psi \r$ defined by \inine\ is given directly in terms
of the Euclidean two-point function \ieleven,
\eqn\itwelve{
\l \psi_\O |\psi_\O \r = \l  \O^E(\tau , \bx )\,  \O^E(-\tau , \bx ) \r =
C_\O \, {1\over (2\tau)^{2\eta}} \ge 0\, ,
}
which therefore requires positivity of $C_\O$ as a consequence of
unitarity.

We now turn to analyze the restrictions which
follow from eqs. \isix\ and \iseven. As a preliminary result, we
need the computation of
the Euclidean three-point function involving the energy momentum tensor,
which is also uniquely determined by conformal invariance \hughone
\eqn\ithirteen{\eqalign{
\l \O^E(x) \, & \O^E(y)\, T^E{}_{\! \alpha\beta}(z) \r \cr
& = -{C_\O\over S_d}\, {d\eta\over d-1} \, {1\over
\big((x-z)^2 (y-z)^2\big )^{\hh d} \big ( (x-y)^2 \big )^{\eta - \hh d} }
\Big ({Z_\alpha Z_\beta \over Z^2}- {1\over d}\de_{\alpha \beta} \Big )\, ,\cr}
}
where $S_d = 2\pi^{\hh d}/\Gamma(\hh d)$ and
\eqn\ifouteen{
Z_\alpha = {(x-z)_\alpha \over (x-z)^2} - {(y-z)_\alpha \over (y-z)^2} \, .
}
The overall coefficient in \ithirteen\ is determined in terms of $C_\O$ by Ward
identities which relate the three-point function \ithirteen\ to the two-point
function given by \ieleven. The relation between amplitudes involving
tensor operators such as $T_{\mu\nu}$ and the associated  Euclidean
correlation functions is given by a matrix\footnote{${}^3$}
{Indices $\alpha,\beta,\gamma,\dots$ for Euclidean correlation functions are
solely subscripts while Minkowski vector indices $\mu,\nu,\dots$  are raised
and lowered with the standard metric $g_{\mu\nu}$, where $g_{00}=-1,\,
g_{ij}=\de_{ij}$.},
$\theta_{\mu\alpha}$, so that
\eqn\ififteen{
\l \psi_\O | T_{\mu\nu}(0)  |\psi_\O \r = \theta_{\mu\alpha} \theta_{\nu\beta}
\l  \O^E(\tau , \bx )\,  \O^E(-\tau , \bx ) T^E{}_{\! \alpha\beta}(0,{\bf 0})\r
\, ,
}
where
\eqn\isixteen{
\theta_{\mu\alpha} = \pmatrix { i& {0}\cr {0} & \de_{ ij}\cr} \, .
}
Hence from \ithirteen\ it is straightforward to see that
\eqn\iseventeen{
\l \psi_\O | n^\mu n^\nu T_{\mu\nu}(0) |\psi_\O \r = \eta \, {C_\O \over S_d}\,
{d-1 + \bn^2 \over d-1} \, {1\over (\tau^2 + \bx^2)^d (2\tau)^{2\eta-d}} \, .
}
Since we assume $\eta>0$ the positivity condition \isix\ gives nothing new in
this case.
Moreover, setting $\bn = {\bf 0}$ and using \iten\ gives
\eqn\ieighteen{
\l \psi_\O | H |\psi_\O \r = 
\int \! \d^{d-1} x \, \l \psi_\O | T_{00}(0)|\psi_\O \r
= 2\eta\, {C_\O\over (2\tau)^{2\eta+1}}
= - \half {\pr \over \pr \tau}\l \psi_\O |\psi_\O \r \, ,
}
which provides an additional check on the overall normalisation in
 \ithirteen.

\newsec{Correlators involving vector currents}

A less trivial example is provided by considering the state formed by the
action of a conserved vector operator, which must have dimension $d-1$, on
the vacuum. The relation of the matrix elements to Euclidean correlation
functions involves applying also the matrix $\theta$, 
defined in \isixteen, to the vector indices and the algebra becomes more 
complicated. On the other hand, the two-point correlation function for the
vector current is still characterised by a single overall parameter in
conformal field theories and there are no anomalous dimensions present in this
case.

As in the previous section, we first construct the state
\eqn\inineteen{
|\psi_V \r = e^{-H\tau} V_\mu(0,\bx) |0\r \psi^\mu \, , \quad \tau >0 \, .
}
We have contracted the $V_\mu$ operator with an external vector
$\psi^\mu$, which can be chosen freely and which allows
positivity conditions to be investigated for all possible
combinations for Lorentz indices with ease. The norm of the state
is directly related to the Euclidean two-point function
\eqn\itwenty{
\l \psi_V |\psi_V \r = \psi^\mu \psi^\nu \theta_{\mu\alpha}
\theta_{\nu\beta}\l  V^E{}_{\!\alpha}(\tau , \bx )\,
V^E{}_{\!\beta}(-\tau , \bx ) \r \, ,
}
and in the conformal limit we have the simple form
\eqn\itwentyone{
\l  V^E{}_{\!\alpha}(x) \, V^E{}_{\!\beta}(0) \r = C_V \, {1\over x^{2(d-1)}}
I_{\alpha \beta}(x) \, , \qquad
I_{\alpha \beta}(x) \equiv \de_{\alpha\beta}-2{x_\alpha x_\beta \over x^2} \, ,
}
where $I_{\alpha \beta}(x)$ represents the action of inversions. From 
\itwentyone\ it is easy to see that
\eqn\itwentytwo{
\l \psi_V |\psi_V \r = C_V\, {1\over (2\tau)^{2(d-1)}} \psi^\mu \psi^\mu \, ,
}
which is manifestly positive so long as $C_V>0$. We should note here that
the initially surprising sum over 
two upper $\mu$ indices, so that there is an effective Euclidean metric, 
is essential to ensure positivity.

We now consider the conditions flowing from the assumption of the
positivity condition,
\eqn\itwentythree{
\l \psi_V | n^\mu n^\nu T_{\mu\nu}(0) |\psi_V \r \ge 0 \, ,
}
since $\l 0 |  T_{\mu\nu}(0) |\psi_V\r = 0$, without restriction on $\psi^\mu$.
The matrix element in
\itwentythree\
is directly related to the Euclidean three-point function
$\l  V^E{}_{\!\gamma}(x) \, V^E{}_{\!\de}(y) \, T^E{}_{\!\alpha\beta}(z)\r$
which in the conformal limit has two possible linearly independent forms
\hughone. As mentioned in the introduction for
simplicity we consider the case when $x,y,z$ are collinear, along the
direction defined by $\he_\alpha =(1,{\bf 0})$, but appendix
B shows that the results obtained for the collinear configuration
are equivalent to the more general case assuming $\bx \ne {\bf 0}$.
With this simplification the three-point
function is restricted to the simple form,
\eqn\itwentyfour{
\l  V^E{}_{\!\gamma}(\hx\he) \, V^E{}_{\!\delta}(\hy\he) \,
T^E{}_{\!\alpha\beta}(\hz\he)\r
= {1 \over |\hx-\hz|^{d}\, |\hy-\hz|^{d}\,
|\hx-\hy|^{d-2}} \, \A_{\gamma\delta\alpha\beta} \, ,
}
where $\A_{\gamma\de\alpha\beta} = \A_{\de\gamma\alpha\beta}
= \A_{\gamma\de\beta\alpha}, \, \A_{\gamma\de\alpha\alpha} = 0$
is an invariant tensor under $O(d-1)$ transformations
leaving $\he$ invariant. Using the notation
$\he_\gamma\he_\de \A_{\gamma\de\alpha\beta} = \A_{\he\he\alpha\beta}$
we may therefore write
\eqn\itwentyfive{
\A_{\he\he mn}= \beta \, \de_{mn} \, , \quad \A_{i\he m\he}= \de \,\de_{im}\, ,
\quad \A_{ijmn} = \rho \,\de_{ij} \de_{mn} + \tau
(\de_{im}\de_{jn} + \de_{in} \de_{jm} ) \, ,
}
with other components zero or determined by the traceless condition on
$\alpha\beta$. The conservation conditions for the vector current and the
energy momentum tensor lead to two conditions
\eqn\itwentysix{
\beta + \rho + d\tau = 0 \, , \qquad 2\de + d\rho + (d+2)\tau =0 \, ,
}
so there are left two independent parameters which may be taken as $\rho,\tau$.
Ward identities give a relation to the coefficient of the two-point function
\eqn\itwentyseven{
C_V = {S_d \over d} \big ( d\rho + (d+2)\tau \big ) \, ,
}
so this combination must be positive.

If in \inineteen\ we set $\bx={\bf 0}$ then using \itwentyfour\ we may
obtain in the conformal
limit
\eqn\itwentyeight{
\l \psi_V | n^\mu n^\nu T_{\mu\nu}(0) |\psi_V \r = {1\over \tau^{2d}
(2\tau)^{d-2}} \psi^\si \psi^\rho  M_{\si\rho} \, , \quad
M_{\si\rho} = n^\mu n^\nu \theta_{\mu\alpha} \theta_{\nu\beta}
\theta_{\si\gamma} \theta_{\rho\de}\A_{\gamma\de\alpha\beta} \, ,
}
where explicitly
\eqn\itwentynine{
M_{\si\rho} =\pmatrix{-(d-1+\bn^2)\beta  & - 2 \delta \, n_j \cr
- 2 \delta \, n_i & \big ( (d-1 +\bn^2)\rho + 2\tau \big ) \de_{ij}
+ 2\tau \, n_i n_j \cr } \, .
}
The positivity condition \itwentythree\ then reduces to the positivity of the
matrix $M$.
For $\bn = {\bf 0}$ this requires
\eqn\ithirty{
- \beta =  \rho + d\tau \ge 0 \, , \quad (d-1)\rho + 2\tau \ge 0 \, ,
}
and it is easy to see that together they give $C_V>0$ using \itwentyseven.
For $\bn$ non zero  it is convenient to write $\psi^\si=(a, b{\hat \bn}+\bv)$
with $\bv{\cdot \bn}=0, \, {\hat \bn} = \bn/|\bn|$ so that
\eqn\ithirtyone{
\psi^\si \psi^\rho  M_{\si\rho} = \big ( (d-1+\bn^2)\rho + 2\tau
\big ) \bv^2 + \V^T \M \V \, ,
}
where
\eqn\ithirtytwo{
\V = \pmatrix{a\cr b\cr} \, \qquad
\M =\pmatrix{-(d-1+\bn^2)\beta & - 2 \delta \, |\bn| \cr
-2  \delta \, |\bn| & (d-1+\bn^2)\rho + 2\tau(1+\bn^2) \cr } \, .
}
If $\bn^2 = 1$ then
\eqn\ithirtythree{
\det \M = (d-2)\big( d^2 \rho + (3d+2) \tau \big ) \tau \, .
}
The positivity requirements then reduce to $(d-1+\bn^2)\rho + 2\tau \ge 0$
and positivity of the eigenvalues of the matrix $\M$, which of course
implies $\det \M \ge 0$. It is not difficult to show that the necessary and
sufficient conditions, for $d> 2$, can be reduced to
\eqn\ithirtyfour{
d \rho + 2 \tau \ge 0 \, , \qquad \tau \ge 0 \, ,
}
which are stronger than \ithirty. It is evident that in this case the
positivity
conditions on the energy momentum tensor lead to requirements which go beyond
what can be found by consideration of the two-point function alone.

It is of course crucial to check whether these conditions are compatible
with known results which means comparing with explicit calculations
for free fields. The two relevant cases are free $n$-component scalar fields,
in which there are conserved currents corresponding to the $SO(n)$ symmetry,
and also Dirac fermions. In both cases for massless fields there is a conserved
traceless energy momentum tensor for general $d$.
Neglecting inessential positive constants previous calculations give
\eqn\ithirtyfive{
\rho_S = d-4 \, ,\quad \tau_S = d \, ; \qquad\qquad \rho_F = 1 \, , \quad
\tau_F = 0 \, .
}
Thus the positivity conditions \ithirtyfour\ or \ithirty\
are met in both these cases.

The above example is indicative   of the general type of
results which may arise from energy conditions in quantum field theory.
A local energy density positivity condition should, if valid, enforce
some constraints on three-point functions involving the energy momentum tensor
which in turn transform
into inequalities for the parameters defining the three point function at
a conformally invariant fixed point.
These constraints may be checked against
known results, which in practice are restricted to free field theories.

\newsec{Correlators involving energy momentum tensors}

We finally consider the case where all correlation functions are
made out of energy momentum tensors. This instance is of particular
interest since some important parameters characterising the theory
show up through properties of this operator, which was the motivation
for this investigation\footnote{${}^4$}{This was first considered
some time ago by Cappelli and Latorre 
\ref\latorre{A. Cappelli and J.I. Latorre, unpublished.} and reported
on in ref. \ref\hughthree{H. Osborn, {\it in} Proceedings of the XXVII
Ahrenshoop International Symposium, DESY 94-053.}}. Furthermore, there
are well-established relations between trace anomaly
and operator product expansion coefficients, which
are computable {\sl via} two and three-point
stress tensor correlation functions
in four dimensions. Our positivity results are potentially relevant 
in both of these areas.

 We therefore consider the state
\eqn\ithirtysix{
|\psi_T \r = e^{-H\tau} T_{\mu\nu}(0,\bx) |0\r \ \psi^{\mu\nu} \, ,
\quad \tau >0 \, .
}
The two-point function of the energy momentum tensor after analytic
continuation to Euclidean space in the conformal limit is given by a simple
generalisation of \itwentyone
\eqn\ithirtyseven{
\l T^E{}_{\!\alpha\beta}(x)\, T^E{}_{\!\gamma\de}(0) \r = C_T\, {1\over x^{2d}}
\Big ( \half \big(I_{\alpha\gamma}(x)I_{\beta\de}(x) + I_{\beta\gamma}(x)
I_{\alpha\de}(x)\big ) - {1\over d} \de_{\alpha\beta}\de_{\gamma\de}\Big ) \, .
}
With this result
\eqn\ithirtyeight{
\l \psi_T |\psi_T \r = \psi^{\si\rho} \psi^{\kappa\lambda}
\theta_{\si\gamma} \theta_{\rho\delta}\theta_{\kappa\ep}\theta_{\lambda\eta}
\l  T^E{}_{\!\gamma\de} (\tau ,\bx ) \, T^E{}_{\!\ep\eta}(-\tau ,\bx ) \r
= C_T \, {1\over (2\tau)^{2d}} \psi^{\si\rho}\psi^{\si\rho} \, ,
}
if we impose $g_{\mu\nu}\psi^{\mu\nu}=0$ or
$\psi^{00}= \psi^{ii}$. Just as with previous examples,
unitarity requires the positivity of $C_T$.

In order to analyze the more subtle condition \isix, in conjunction with
\iseven, we now set $\bx={\bf 0}$ in the definition
of the state $|\psi_T \r$ so that  it is  easy to see that
\eqn\ithirtynine{
\l 0 |n^\mu n^\nu  T_{\mu\nu}(0) |\psi_T \r = C_T \, { 1\over \tau^{2d}} \,
n^\mu n^\nu \psi^{\mu\nu} \, .
}

With $\bx={\bf 0}$ the analysis of the matrix element
$\l \psi_T | n^\mu n^\nu  T_{\mu\nu}(0) |\psi_T \r$ can be reduced to the
collinear Euclidean three-point function which can be written simply as
\eqn\ifourty{
\l T^E{}_{\!\alpha\beta}(\hx \he) \,T^E{}_{\!\gamma\de} (\hy \he) \,
T^E{}_{\!\ep\eta} (\hz \he) \r
= {1\over |\hx-\hy|^{d}\, |\hx-\hz|^{d}\,
|\hy-\hz|^{d}} \, \A_{\alpha\beta\gamma\de\ep\eta} \, ,
}
where $\A_{\alpha\beta\gamma\de\ep\eta}$ is an invariant tensor under $O(d-1)$
rotations preserving $\he$ symmetric and is traceless for
each pair of indices $\alpha\beta, \, \gamma\de, \, \ep\eta$ and also
symmetric under interchange of each pair. The essential components, with
a similar notation to \itwentyfive, are
\eqn\ifourtyone{ \eqalign {
\A_{ijk\he m\he} = {}& \rho \,\de_{ij}\de_{km} + \tau
(\de_{ik}\de_{jm} + \de_{im} \de_{jk} ) \, ,\cr
 \A_{ijk\ell mn} = {}&r \,\de_{ij}\de_{k\ell}\de_{mn}\cr
{}& + s\bigl ( \de_{ij}(\de_{km}\de_{\ell n} + \de_{kn} \de_{\ell m} )
+ \de_{kl}(\de_{im}\de_{jn} + \de_{in} \de_{jm} )
+ \de_{mn}(\de_{ik}\de_{j\ell} + \de_{i\ell} \de_{jk} )\bigl) \cr
{}& + t \big (\de_{ik} \de_{jm} \de_{\ell n} + i\leftrightarrow j,
k\leftrightarrow \ell , m \leftrightarrow n \big ) \, , \cr}
}
with others determined by the symmetry and traceless conditions. The
conservation equations give two conditions
\eqn\ifourtytwo{
2\tau + dr + (d+4)s + 2t =  0 \, , \quad
2\rho - dr + (d-2)s + 2(d+4) t = 0 \, ,
}
so there remain three independent parameters which may be taken as $r,s,t$.
One linear combination is determined in terms of the coefficient of the
two-point function in \ithirtyseven\ by a Ward identity
\eqn\ifourtythree{ \eqalign {
C_T = {}& {2S_d\over d(d-1)(d+2)} \Big ( d \beta -(d-2)(d+1)\ep + 2(d-1)\gamma
\Big ) \cr
= {}& {4S_d\over d(d+2)} \Big ( dr + (d^2+2d+4)s + (d^2+5d+2)t \Big ) \, ,\cr}
}
where we have introduced the alternative variables $\beta,\epsilon,\gamma$
for later convenience by
\eqn\ifourtyfour{
\beta = (d-1)^2 r + 6(d-1) s + 8t \, , \quad -\ep = (d-1)s+4t \, , \quad
\gamma = -(d-1)\rho-2\tau \, .
}

With these results we may obtain the analogous expression to
 \itwentyeight\
 in this case
\eqn\ifourtyfive{
\l \psi_T |  n^\mu n^\nu T_{\mu\nu}(0) |\psi_T \r
= {1\over \tau^{2d}(2\tau)^d }\, \psi^{\si\rho}\psi^{\kappa\lambda}
M_{\si\rho,\kappa\lambda} \, ,
}
where $M_{\si\rho,\kappa\lambda}=M_{\kappa\lambda,\si\rho}$, which forms a
$\half (d+2)(d-1) \times \half (d+2)(d-1)$ matrix, can be expressed in terms of
$\A_{\mu\nu\si\rho\kappa\lambda}$ giving
\eqn\ifourtysix{ \eqalign {
M_{0m,0n} = {}& \gamma\, \de_{mn} - \rho \bn^2 \de_{mn} - 2\tau\, n_m n_n\, ,
\cr
M_{0m,k\ell} = {}& - 2 \big ( \rho \,\de_{k\ell} n_m
+ \tau (\de_{mk}n_\ell+ \de_{m\ell}n_k)\big ) \, , \cr
M_{ij,k\ell} = {}& {\beta \over d-1} \de_{ij} \de_{k\ell}
- \ep \Big ( \de_{ik}\de_{j\ell}
+ \de_{i\ell}\de_{jk} - {2\over d-1} \de_{ij} \de_{k\ell} \Big ) \cr
{}& + r \,\de_{ij} \de_{k\ell} \bn^2
+ s(\de_{ik}\de_{j\ell}+\de_{i\ell}\de_{jk})
\bn^2 +2s (\de_{ij}n_k n_\ell + \de_{k\ell} n_i n_j ) \cr
{}& + 2t ( n_i n_k \de_{j\ell} + i\leftrightarrow j, k\leftrightarrow \ell)\, ,
\cr}
}
with other components determined by symmetry and conditions such as
$M_{ij,00} = M_{ij,kk}$, giving for instance $M_{00,00}=(d-1+\bn^2)\beta$.

Our positivity conditions then reduce to
\eqn\ifourtyseven{
\psi^{\si\rho}\psi^{\kappa\lambda} M_{\si\rho,\kappa\lambda} \ge 0 \quad
\hbox{if} \quad n^\mu n^\nu \psi^{\mu\nu} = 0 \, .
}
We  analyse  this condition in two stages. First we consider
the case where $\bn = {\bf 0}$. It is easy to see that \ifourtyseven\ and
 tracelessness
of $\psi$ yield $\psi^{00}=\psi^{ii}=0$. Moreover,  a short computation
shows that our proposed  positivity conditions reduce to
\eqn\ifourtyeight{
\quad \gamma  \ge 0 \, , \quad -\ep \ge 0 \, ,
}
although positivity of the full matrix $M_{\si\rho,\kappa\lambda}$ would lead
to $\beta \ge 0$ as well.

More generally, we consider the case $\bn\not={\bf 0}$. To disentangle
the complete set of independent positivity restrictions,
 we  decompose $\psi^{\si\rho}$ in the form
\eqn\ifourtynine{\eqalign{
\psi^{\si\rho}={}&\pmatrix{f+(d-1)g& \half w_n +
\half a \hv_n + \half e \hn_n \cr
\half w_m + \half a\hv_m + \half e \hn_m & u_{ij} +
\half b(\hn_i \hv_j + \hn_j \hv_i)
+ f \hn_i \hn_j + g \de_{ij} \cr} \, , \cr
& \bn {\cdot \bw} = \bw {\cdot \hat\bv}= \bn {\cdot \hat\bv} =0 \, ,
\quad u_{ij} = u_{ji} \, , \ u_{ii}=0 \, , \
u_{ij}\hn_j = 0 \, , \cr}
}
so that
\eqn\ifivety{
\psi^{\si\rho}\psi^{\si\rho}= 2(f+\half d g)^2+ \half d(d-2)g^2
+ \half (e^2 + a^2 + b^2) + \half \bw^2 + u_{ij}u_{ij} \, ,}
and
\eqn\ifivetya{ 
n^\si n^\rho \psi^{\si\rho} = (1+\bn^2)f + (d-1+\bn^2)g + |\bn| e \, .}
Using \ifourtynine\ we may find
\eqn\ifivetyone{
\psi^{\si\rho}\psi^{\alpha\beta} M_{\si\rho,\alpha\beta} = 2(-\ep+s\bn^2)
u_{ij}u_{ij} + (\gamma-\rho \bn^2) \bw^2 +
\V_2{}^{\!T} \M_2 \V_2 + \V_3{}^{\!T} \M_3 \V_3 \, ,
}
where
\eqn\ifivetytwo{
\V_2=\pmatrix{a\cr b\cr}\, , \quad
\M_2 = \pmatrix{\gamma - \rho \bn^2 & - 2 \tau |\bn| \cr
-2 \tau |\bn| & - \ep + (s+2t)\bn^2 \cr} \, .
}
and, assuming
\eqn\ifivetythree{
\V_3 = \pmatrix{e\cr f\cr g \cr} \, ,
}
then
\eqn\ifivetyfour{ \!\!\!\!\!\!
\M_3 =  \pmatrix{\gamma-(\rho+2\tau)\bn^2& -2(d\rho+4\tau)|\bn| &
2 d \gamma|\bn| \cr -2(d\rho+4\tau) |\bn| & X &
{d^2\over d-1}(d-1+\bn^2)\beta -2d{d-2\over d-1}\ep \bn^2
\cr 2d\gamma |\bn|
& {d^2\over d-1}(d-1+\bn^2)\beta -2d{d-2\over d-1}\ep \bn^2
& d^2(d-1+\bn^2)\beta  \cr}
}
where
$$
X =  {1\over d-1}\big( d^2 \beta - 2(d-2)\ep \big) +
{1\over d-1}\big( (d+1) \beta - 4(d-2)\ep \big)\bn^2 + (r+6s+8t)\bn^2
\, .
$$
The positivity condition \ifourtyseven\ as applied to \ifivetyone\
 requires positivity of $-\ep+s\bn^2$, $\gamma-\rho \bn^2 $
as well as  of the matrices $\M_2$ and $\V_3{}^{\! T}\M_3\V_3$, where
the vector $\V_3$ is constrained by the condition 
$n^\si n^\rho \psi^{\si\rho}=0$, for all $|\bn|$ such that $0\le |\bn| \le 1$.
In particular, positivity of $\M_2$ for
$\bn = \bf 0$ leads to our previous result \ifourtyeight. Setting $|\bn|=1$
the positivity conditions from $\psi^{\si\rho}\psi^{\alpha\beta} 
M_{\si\rho,\alpha\beta}\ge 0$, excluding those involving the matrix 
$\M_3$, reduce to the linear relations
\eqn\extra{
-\ep+s = ds+4t \ge 0 \, , \qquad \gamma-\rho = -d\rho - 2\tau \ge 0 \, , }
as well as  the nonlinear condition
\eqn\extraQ{
Q \equiv \det \M_2 \big |_{|\bn|=1} =
\det \pmatrix{ -d\rho - 2\tau & -2\tau\cr - 2\tau & ds + 6t} \ge 0 \, . }

Only the conditions arising from $\M_3$ are sensitive to the  requirement
$n^\si n^\rho \psi^{\si\rho}=0$ since from \ifivetya\ this constrains only
$\V_3$. For $|\bn|=0$ then we may eliminate $f$ in terms of $g$, $f=-(d-1)g$,
 so that the previous results
\ifourtyeight\ are recovered. If we take $|\bn| =1$ then we choose to
eliminate $e$ in favour of $f$ and $g$ and the matrix 
acting on $\pmatrix{f\cr g\cr}$ becomes
\eqn\ifivetyfive{
{\overline \M}= (d-2)\pmatrix{(d+4)(dr + 4s -4t)&d(d+3)(dr + 4s -4t)\cr
d(d+3)(dr + 4s -4t)& \half d^2 \big( d(2d+3)r + 3(3d+4)s - 6(d+2)t\big)\cr} \,
.
}
If $d>2$ the positivity of ${\overline \M}$ reduces to
\eqn\ifivetysix{
I \equiv dr + 4s -4t = {d^2\beta-4(d-1)\gamma-2(d-2)\epsilon \over
(d-1)(d-2)(d+3)} \ge 0 \, ,
}
and also, by considering $\det {\overline \M}$,
\eqn\ifivetyseven{
J \equiv - d(d+6)r + (d^2-24)s + 2(d^2 + 6d + 12) t \ge 0 \, .
}
Note that, from \ifourtythree,
\eqn\ifivetyeight{
C_T = {2S_d \over d^2(d+2)}\Big ( (d-2)(d+3)I + (d+2)\big ( 2\gamma
- (d-2) \epsilon \big ) \Big ) \, ,
}
which by virtue of \ifourtyeight\ and \ifivetysix\
 is manifestly positive as is necessary
for unitarity.

The detailed analysis of the mutual interdependence of the constraints
obtained in this case is obviously more complicated than previously. 
It is perhaps of interest first to consider the case $d=3$ separately since, as
shown in \hughone, there are then only two independent parameters which
may be taken as
\eqn\dthree{
u = r-4t \, , \qquad v= s+ 2t\, .}
Note that $C_T = {16\over 15}\pi(3u+19v)$ and for the quantities appearing
in the various inequalities (if $d=3$ then $u_{ij}$ in \ifourtynine\ is absent
so that there is no condition for $-\ep+s$),
\eqn\dthreea{
-\ep={\ts{1\over 8}}\gamma=v \, , \quad I= 3u+4v \, , \quad J=-3(9u+5v)
\, , \quad \gamma -\rho = \half (17v-3u) \, .}
The inequalities lead to $-{5\over 9}\ge u/v \ge -{4\over 3}$.

When $d=4$, which is of primary interest, three parameters are necessary
but the positivity conditions are homogeneous so that they essentially
constrain their ratios. The relationship between different conditions is
most easily visualised in terms of the two-dimensional plot given by Fig. 1.
{\midinsert
\hfil \epsfysize=0.4\vsize
\epsfbox{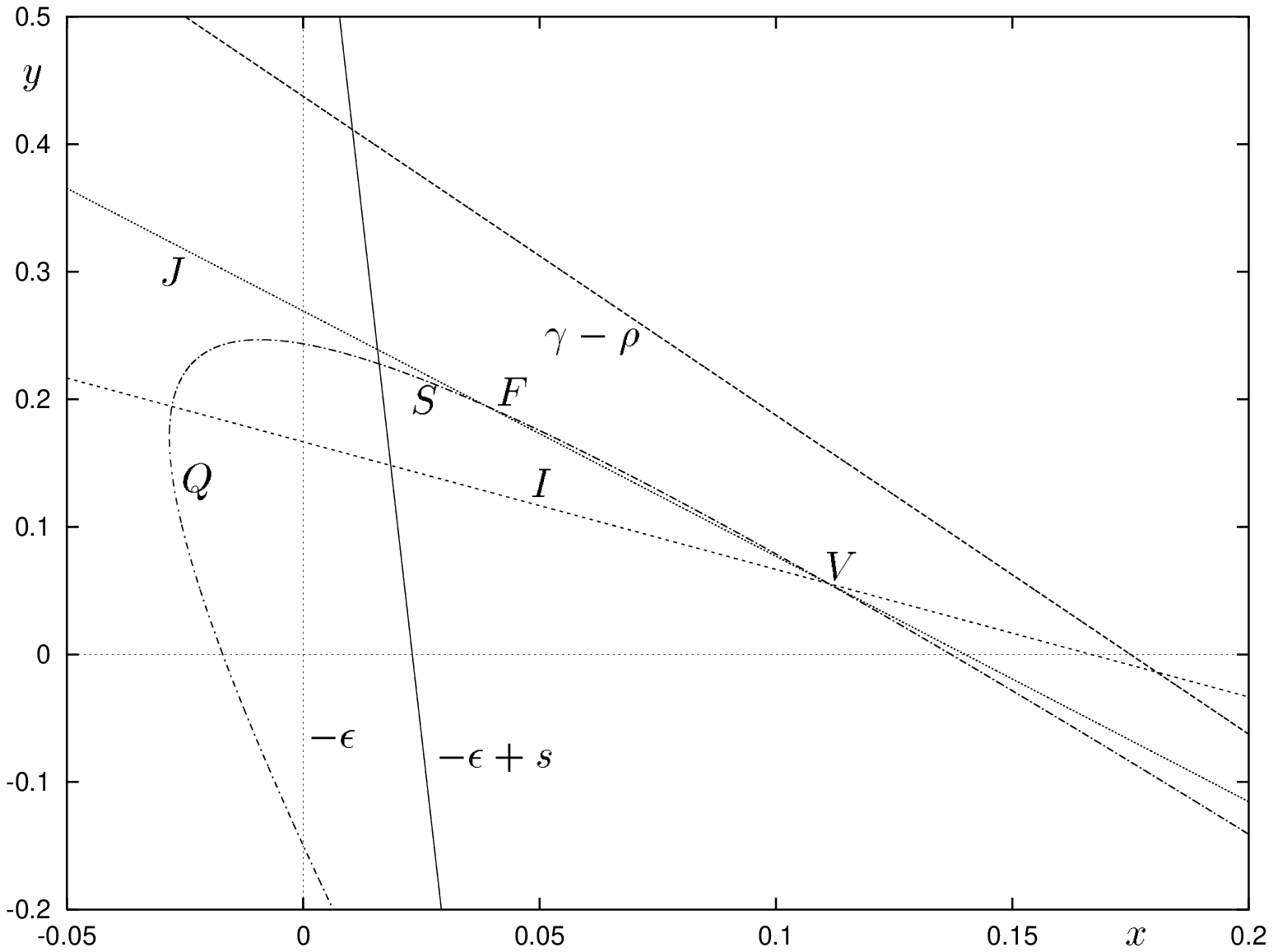} \hfil
\vskip -12pt
{\eightpoint{
\parindent 1.5cm

{\narrower\smallskip\noindent 
Fig. 1 Boundaries of various positivity inequalities for $d=4$ on graph with
coordinates $y= \beta/(\beta-5\epsilon+3\gamma)$ and $x=-\epsilon/
(\beta-5\epsilon+3\gamma)$, where $\beta-5\epsilon+3\gamma \propto
C_T >0$. Each boundary is labelled by the corresponding quantity which
is required to be positive on the side for which it is positive. The allowed
region satisfied by all constraints is the central triangular region
with one corner cut off by the $Q=0$ curve. The points labelled by $V$, $F$, 
which lie on
the boundary of the positivity region, and $S$ correspond to free vector,
fermion and scalar fields.\smallskip}

\narrower}}
\endinsert}

The relevance and consistency of our results may be checked by investigating 
the cases of the essentially trivial conformal field theories
formed by free scalars, fermions, for general $d$, and also vector fields
when $d=4$. Neglecting inessential
positive constants we have from \hughone
\eqn\ifivetyseven{ \eqalign {
r_S = {}& d^3 +28 d -16 \, , \quad s_S = d(d^2 - 8d +4) \, , \quad t_S =d^3 \,
,
\cr
r_F = {} & - 4 \, , \qquad s_F = d \, , \qquad t_F =0 \, ,\cr
d={}& 4 \, , \qquad r_V = -3 \, , \qquad s_V = 2 \, , \qquad t_V = -1 \, . \cr}
}
We have used these results to check all the various inequalities, for $d=4$
they give the points shown in Fig. 1.
For general free conformal field theories for $-\ep+s$,
which featured in \extra, we find
\eqn\ifivetysevena{
(-\ep+s)_S = d^2(d-2)^2 \, , \qquad (-\ep+s)_F = d^2 \, , 
\qquad (-\ep+s)_V=4 \, ,}
so this inequality is respected in each case and also with the definition in 
\extraQ\
\eqn\ifivetyeight{
Q_{S} = d^4(d+2)(d-1)(d-2)^3 \, , \quad
Q_{F} = \half d^4(d+2) \, , \quad Q_{V} = 0 \, .
}
For the matrix $\M_3$ \ifivetyfour\ for $|\bn|=1$ we find
\eqn\ifivetynine{ \eqalign{
\det \M_{3,S} = {}& 2d^6 (d+2)(d-1)(d-2)^4(d^5+d^4-10d^3-4d^2-24d+32) \, , \cr
& \det \M_{3,F} =  \det \M_{3,V} = 0 \, . \cr}
}
It is easy to see that $\det \M_{3,S} < 0$ when $d=3$.
Nevertheless from the expressions for $I,J$ in the more restricted conditions
\ifivetysix\ and \ifivetyseven\  we find
\eqn\isixty{ \eqalign{
I_S = {}& d^2(d+2)(d-2) \, , \qquad I_F = I_V = 0 \, , \cr
J_{S} = {}& 2d^2(d-2)(d^2+d-10) \, , \quad J_F = d^2(d+4) \,
 , \quad J_V=0\, ,
\cr}
}
which are positive for $d=3$. This result shows how the  
restriction on the states $| \psi \r$ to the subspace defined by \iseven\ 
is sufficient to maintain the fulfilment of the positivity conditions
in free scalar theories in three dimensions. In terms of the variables in
\dthree\ we have $(u/v)_S = - {13\over 21}, \, (u/v)_F= -{4\over 3}$ which
lie in the required positivity range.

However $J_S$ in \isixty\ is no longer positive for $d$ close to two although
when $d=2$ exactly this problem is absent since there is just one
unique expression for the conformally invariant
energy momentum tensor three-point function whose
coefficient is determined by the value of $C_T$, which is proportional to
the Virasoro central charge $c$. Nevertheless the lack of positivity
of $J_S$ if $2 < d < \half (\sqrt{41}-1)$ suggests, following
the guidelines exemplified by the quantum mechanical analogues discussed
in Appendix A, that further restrictions are necessary.  

\newsec{A refined energy density positivity postulate}

The problems noted above for scalar field theories are confined to
the matrix ${\overline \M}$ defined in \ifivetyfive. As mentioned above
it is natural to introduce additional constraints on
the state $|\psi_T\r$. Any further such conditions will require some
knowledge
of the detailed dynamics and will therefore be model dependent to this extent. 
As a possible extra constraint on the states $|\psi\r$, which may be
natural in the context of conformal field theories, we postulate,
as well as \iseven,
\eqn\cone{
\l \psi_\O | n^\mu n^\nu T_{\mu\nu}(0) | \psi \r = 0 \, ,
}
where $|\psi_\O\r$ is constructed as in \inine\ in terms of a scalar field $\O$
appearing in the theory.  It is useful to note that
\eqn\ctwo{
\l \O^E(x) \O' {}^E(y) T^E{}_{\! \alpha\beta}(z) \r = 0 \ \ \hbox{if} \ \
\eta \ne \eta' \, , \quad \l \O^E(x) V^E{}_{\! \gamma}(y)
T^E{}_{\! \alpha\beta}(z) \r = 0 \, ,
}
and hence this extra condition will not affect the positivity condition 
\itwentythree\
for states formed by a vector operator which led to \ithirty\ or 
\ithirtyfour but will impose only restrictions on any conditions
arising from the matrix ${\overline \M}$ as required.

To apply the condition \cone\ to the states $|\psi_T\r$ we make use of results
for the Euclidean three-point function $\l \O^E(x) T^E{}_{\! \gamma\de}(y)
T^E{}_{\! \alpha\beta}(z)\r$ which in the conformal limit has only one
linearly independent form. In the collinear configuration we have
\eqn\cthree{
\l \O^E(\hx\he) T^E{}_{\! \gamma\de}(\hy\he) T^E{}_{\! \alpha\beta}(\hz\he)\r
= { 1 \over |\hx - \hy |^\eta |\hx -\hz|^\eta |\hy -\hz|^{2d-\eta}} \,
\B_{\gamma\de\alpha\de} \, ,
}
where $\B_{\gamma\de\alpha\de}$ is an $O(d-1)$ invariant tensor given by
(note that here $\de, \, \epsilon , \, \gamma$ are new parameters unrelated
to those of the previous section)
\eqn\cfour{
\B_{ijk\ell} = \de \, \de_{ij} \de_{ k\ell} + \ep (
\de_{ik}\de_{j\ell} + \de_{i\ell} \de_{jk} ) \, , \quad
\B_{i\he k \he} = \gamma \, \de_{ik}\, ,
}
with other components determined by symmetry and tracelessness. Conservation
of the energy momentum tensor gives two conditions
\eqn\cfive{
2\gamma = - (d-\eta)\big ( (d-1)\de + 2 \ep \big ) \, , \qquad
(d-\eta) \gamma = - d\,\de - (d+2) \ep \, ,
}
which may be solved, up to an inessential overall constant, by taking
\eqn\csix{
\ep = \half (d-\eta)^2 - {d\over d-1} \, , \quad
\de = {1\over d-1}\big ( -(d-\eta)^2 + d+2 \big )  \, , \quad
\gamma = - \half (d-\eta){(d-2)(d+1)\over d-1} \, .
}
We can now write, if $\bx = {\bf 0}$ so that using \cthree\ is 
valid,\footnote{${}^5$}{In general the matrix element may be written as
$\l \psi_\O(\bx',\tau') | n^\mu n^\nu T_{\mu\nu}(0) | \psi_T(\bx,\tau) \r$ but
only if $\bx'=\bx, \tau'=\tau$ can conformal invariance be used to transform
this to the collinear configuration, as in appendix B, without effectively
transforming $n^\mu$. It is perhaps  simpler
to consider just the state $|\O\r= \lim_{\tau\to\infty} (2\tau)^\eta
e^{-H\tau} \O(0,\bx) |0\r$, which has norm from \itwelve\ 
$\l \O | \O \r = C_\O$, and impose
$\l \O | n^\mu n^\nu T_{\mu\nu}(0) | \psi \r = 0$ instead of \cone\ as an
additional condition on the states $| \psi \r $ to which the positivity
condition is applied. In this
situation we can restrict our attention to the collinear case.}
\eqn\cseven{
\l \psi_\O | n^\mu n^\nu T_{\mu\nu}(0) | \psi_T \r =
{1\over \tau^{2d}(2\tau)^\eta}\, n^\mu n^\nu \psi^{\si \rho}
\theta_{\mu\gamma} \theta_{\nu\de}\theta_{\si\alpha}\theta_{\rho\beta}
\B_{\gamma\de\alpha\de} \, .
}
Using \cfour\ we can write, with $g_{\mu\nu}\psi^{\mu\nu}=0$,
\eqn\ceight{
\eqalign{
n^\mu n^\nu \psi^{\si \rho} &
\theta_{\mu\gamma} \theta_{\nu\de}\theta_{\si\alpha}\theta_{\rho\beta}
\B_{\gamma\de\alpha\de}\cr
& =  \de\, d(d-1+\bn^2) \psi^{ii}+ 2\ep \big (
(d+\bn^2) \psi^{ii} + n^i n^j \psi^{ij} \big ) - 4\gamma \, \psi^{0i}n^i \cr
& = \eta\, {d-2 \over d-1} \Big ( \big ( (\eta-d)\bn^2 + d+1+\bn^2 \big ) f
+ (d+1)(d-1+\bn^2) g \Big ) \, , \cr}
}
where we have used \ifourtynine\ and \csix\ as well as 
$n^\si n^\rho \psi^{\si \rho}=0$
to eliminate $e$ if $\bn \ne {\bf 0}$. If $\bn = {\bf 0}$ then this vanishes
if $\psi^{ii}=0$ which is the same condition as that already obtained from
$\l 0 | n^\mu n^\nu T_{\mu\nu}(0) | \psi_T \r = 0$.

In order to analyse the consequences of the extra constraint \cone\ for
$|\psi\r \to |\psi_T\r$ we set $|\bn| = 1$ and then, instead of positivity
of the matrix ${\overline \M}$, we require now only that
\eqn\cnine{
\V_{\eta}{}^{\! T} \,{\overline \M} \,\V_\eta \ge 0 \, ,\qquad
\V_\eta = \pmatrix { -d(d+1)\cr \eta+2\cr} \, .
}
If $\eta$ were a free parameter this would reduce to positivity of
${\overline \M}$ again but in any conformal theory the spectrum of
scale dimensions $\eta$ is bounded below by some positive number. For
free scalars, which produced the earlier difficulties, the lowest dimension
scalar operator which could contribute here is $\phi^2$ having dimension
$d-2$. Inserting the results \ifivetyseven\ we find
\eqn\cten{
{\overline \M}_S = (d-2)^2 d^2 \pmatrix { (d+2)(d+4) & d(d+2)(d+3) \cr
d(d+2)(d+3) & d^2(d^2+5d+2) \cr}
}
so that
\eqn\celeven{
\V_{d-2} \,{\overline \M}_S \,\V_{d-2} = (d-2)^2 d^4 (d^3+d^2+10d+8) \, ,
}
which is always positive. For free fermions or, if $d=4$, free vectors
\eqn\ctwelve{
{\overline \M}_F = \pmatrix{0&0\cr0& \half d^4(d-2)\cr} \, , \qquad
{\overline \M}_V = 0 \, ,
}
which satisfy the previous positivity conditions without further restriction.
In these theories the lowest dimension scalar operator which may be relevant
has a dimension of at least $\eta=d$.

\newsec{Constraints on the trace anomaly coefficients}

At this point we may summarise our main results and try to exploit them
in some related issues that go beyond flat space correlation functions.
In curved space-times negative energy densities appear to play an
essential role in such effects as Hawking radiation and are also connected
with lack of stability of time independent vacuum states.
Conversely, some requirement of positive
energy density might be desired in order to ensure a stable vacuum and to
avoid seeming pathologies like causality violations associated with
wormholes.  We have, thus, tried to resurrect energy conditions in quantum
field theory by postulating  a restricted version of the weak energy
condition, here initially for 
flat space-time as a first step towards finding possible
generalisations of classical conditions.
To circumvent the simplest known counterexamples of violation of such kinds
of positivity conditions we have looked for inspiration in two
quantum mechanical models. This leads us to consider the linear conditions,
defining a restricted subspace,
\eqn\vone{
\l \psi | T | \psi \r \ge 0 \quad
{\rm where}\quad \l \phi_i | n^\mu n^\nu T_{\mu\nu} | \psi \r = 0 \, ,}
for some suitable natural set of states $|\phi_i\rangle$, on which positivity
may then be imposed. It is of course essential
that the subspace be defined in a not too model dependent fashion, although
even if the number of necessary conditions becomes infinite this does not
necessarily lead to them having no content.
The constraints imposed by a simple version of
this postulate which we have investigated
in bosonic, fermionic and vector free field theories, for a
restricted set of states $|\psi\rangle$,  have not led to any
apparent violation but also have given conditions with
a non trivial content.

An important corollary of the above results, and one of our main motivations,
is the generation of constraints
on the coefficients of the trace anomaly in curved space-time backgrounds.
As is well known the expectation value of the trace of the energy
momentum tensor is then non zero even if it vanishes as an operator on
flat space-time, as required in conformal field theories. For $d=4$ the
trace anomaly is formed from dimension 4 scalars constructed from the
Riemann tensor and has the general form,
\eqn\vtwo{g^{\mu\nu}\l T_{\mu\nu} \r_g = -\beta_a F -\beta_b G + h
\nabla^2 R\, ,}
where $F=C^{\alpha\beta\gamma\delta}C_{\alpha\beta\gamma\delta}$ is
the square of the Weyl tensor, $G={1\over 4} \varepsilon^{\mu\nu\rho\sigma}
\varepsilon_{\alpha\beta\gamma\delta}R^{\alpha\beta}{}_{\mu\nu}
R^{\gamma\delta}{}_{\sigma\rho}$ is the Euler density and
the third term is a total derivative. The coefficient $h$ is in fact arbitrary
since it can be altered by adding a local term to the effective action
so we concentrate only on $\beta_a$ and $\beta_b$.

The relation between the coefficient of the Weyl term
and the overall coefficient of the two-point function for the energy momentum 
tensor is well known \jil,
\eqn\vthree{-\beta_a = {\pi^2\over 640} C_T = {\pi^4\over 5760}(2\beta-5\epsilon
+3\gamma) \ge 0\, ,}
using the result \ifourtythree\ for $d=4$.
This term in the trace anomaly is therefore related to the $c$-number term
appearing in the operator product expansion for two energy momentum
tensors  and may be regarded as measuring the reaction 
of the field theory to a shear transformation. The positivity of $-\beta_a$
is thus required by the positivity of $C_T$ which
is rooted in the basic unitarity properties of the Hilbert space, as
illustrated directly by \ithirtyeight.

The trace anomaly constrains the form of the expectation value of the energy
momentum tensor on curved space-times, although it does not essentially
determine it \refs{\fulling,\Visser}
unlike the case in two dimensions. For solutions of Einstein equations $F=G$ 
so the anomaly depends only on $\beta_a+\beta_b$ which has opposite signs 
for different free field theories and hence there are no obvious positivity 
constraints for the anomaly coefficients, 
such as might be found by considering the flux of Hawking 
radiation, to be obtained by analysis of $\l T_{\mu\nu} \r_g$ in applications
to general relativity in four dimensions.

On the other hand, the coefficient $\beta_b$  may play a
significant role in flat space quantum field theories. In many respects it
is the direct analogue of the Virasoro central charge $c$ for two dimensional
theories.
Its analysis is more involved and requires more sophistication than $\beta_a$
since its footprint in a flat space quantum field theory 
lies in three-point correlators of the energy momentum tensor and, 
consequently, its presumptive positivity can only be derived in terms
of some positivity restriction on the energy momentum tensor such as we
have postulated. On conformally flat spaces $F=0$ and then $\int \! \d^4 x
\sqrt g g^{\mu\nu}\l T_{\mu\nu} \r_g $ is proportional to $\beta_b$, the
integral of the Euler density is a topological invariant.
Cardy \cardy\ suggested that $\beta_b$ might be a candidate for proving
a $c$-theorem in four dimensions. A significant
step towards this aim would be to demonstrate its positivity, at least
at renormalisation group fixed points of interacting quantum field theories.

The crucial question is therefore whether the results in section 4 and 5 can 
lead to the positivity of $\beta_b$. Using the work of 
\refs{\hughone,\hughtwo}, we can relate
this coefficient to the parameters in  energy momentum tensor three-point
function as follows
\eqn\vfour{\eqalign{
\beta_b&={\pi^4\over 5760} (4 r + 48 s + 53 t)\cr
       &={\pi^4\over 5760}{1\over 6} ( 2 \beta -95 \epsilon-3
\gamma)\cr}\, .}
In order to discuss the issue of positivity of $\beta_b$ in our framework
there are several alternatives. From the conditions for $\bn = {\bf 0}$
which are given in \ifourtyeight, supplemented by the positivity of $C_T$
which is given in \ifourtythree, we cannot deduce $\beta_b>0$. The constraints
in \extra, \ifivetysix\ and \ifivetyseven\ are stronger. When $d=4$ we have
\eqn\vfive{
-\epsilon+s = {\ts{2\over 105}}(-2\beta-115\epsilon-9\gamma)\, , \quad
 \gamma-\rho = {\ts{1\over 15}}(-2\beta+5 \epsilon+21\gamma) \, , \quad
I = {\ts{2\over 21}}(4\beta - \epsilon - 3\gamma) \, , }
and since
\eqn\vsix{ 
2\beta - 95 \epsilon - 3\gamma 
=  {\ts{10\over63}}(4\beta - \epsilon - 3\gamma)
+ {\ts{16\over21}} (-2\beta-115\epsilon-9\gamma) + {\ts{13\over9}}
(2\beta - 5\epsilon + 3\gamma) \, , }
we evidently have from \vfour\ that a variety of combinations of inequalities
imply $\beta_b>0$ as a corollary of the
version of the weak energy condition postulated in this paper. In terms of
Fig. 1, $\beta_b>0$ corresponds to $y>{1\over 4}(1-100x)$ which is easily seen
to be satisfied by the allowed positivity region. 

The fact that the conditions leading to this result fail at the
peculiar value of $d$ discussed in section 4 for scalar field theories
is perhaps indicative
that further restrictions are necessary. If we drop all conditions
arising from the matrix $\M_3$ and impose just that $C_T$, which is
given by \ifourtythree, is positive as an additional condition
we cannot obtain $\beta_b>0$. 
One more inequality which must involve
$\M_3$ directly is necessary. From section 6 we may consider applying the
extra condition \cone\ for some $\eta$ which still leads to some extra
conditions. As an illustration for a free scalar theory we may
take $\eta=d-2$ and get the constraint
\eqn\vseven{8\beta+5\epsilon+\gamma\ge 0\, .}
This supplementary equation is now sufficient to find a manifestly
positive expression for $\beta_b$, {\sl e.g.} for instance
\eqn\veight{
2\beta - 95 \epsilon - 3\gamma ={\ts{3\over 8}} 
(2\beta - 5\epsilon + 3\gamma)+{\ts{5\over 6}}
(-2\beta-115\epsilon-9\gamma)
+{\ts{17\over120}}(-2\beta+5 \epsilon+21\gamma)+{\ts{2\over 5}}
(8\beta+5\epsilon+\gamma)\ge 0\, .}

In conclusion it is perhaps worth noting that our proposed energy conditions,
while speculative to an extent, are susceptible of further tests. In some
instances the inequalities such as \ithirty, \ithirtyfour\  or
\ifourtyeight, \extra, \extraQ\ and \ifivetysix\ are such that free field
theories lie on the boundary. In such cases it is possible that perturbative
calculations for interacting theories will verify whether the inequalities
are still satisfied. Nevertheless it is perhaps remarkable that, as
exhibited in Fig. 1, that the various inequalities severely constrain the
parameters of the general energy momentum tensor three point function in
a consistent fashion which is compatible with the results of free field
theory and in a fashion which implies $\beta_b>0$.
\bigskip
\noindent{\bf Acknowledgements}
\medskip
We are grateful to the British Council and Spanish MEC 
for financial support allowing for travel to our respective institutions,
also CICYT under contract AEN95-0590, CIRIT under contract GRQ93-1047.
We are also very pleased to acknowledge valuable discussions with Rolf
Tarrach.

\vfill\eject

\appendix{A}{}

As an illustration that our positivity condition makes sense in at least
some circumstances we consider two elementary
standard quantum mechanical examples.

First we consider
the following operator
\eqn\aone{
T= \cosh 2\phi \, a^\dagger a + \half \sinh 2\phi\, ( a^2 + a^{\dagger 2} ) \,
,
}
where $a,a^\dagger$ are conventional annihilation, creation operators. The
expression for $T$ is obviously modelled on the expected form for free field
theory. As is well known $T$ can be diagonalised by a Bogoliubov transformation
\eqn\atwo{
T = a'{}^\dagger a' - \sinh^2 \phi \, ,
}
where
\eqn\athree{
a' = U a U^{-1} = \cosh \phi \, a + \sinh \phi \, a^\dagger \, ,
}
with $U$ a unitary operator given by
\eqn\afour{
U = e^{{1\over 2}\phi (a^2 - a^{\dagger 2})} = (\cosh \phi)^{-{1\over 2}}
e^{-{1\over 2}\tanh \phi \, a^{\dagger 2}} e^{\rho a^\dagger a}
e^{{1\over 2}\tanh \phi \, a^2} \, , \quad \rho = - \ln \cosh \phi \, .
}

For the standard Fock space vacuum $|0\r$, satisfying $a|0\r=0$, it is evident
that the conditions in \ione\ are met if $\phi\ne0$.
For a general state $|\psi\r$ in the Fock space
\eqn\afive{
\l \psi |T|\psi \r = \sum_{n=0} a_n{}^{\! 2} \big ( n - \sinh^2 \phi \big )
\ \ \hbox{for} \ \ |\psi \r = \sum_{n=0} a_n U |n\r \, .
}
Clearly we can choose the state $|\psi \r$ such that
$\l \psi |T|\psi \r<0$ without difficulty
since the $n=0$ term in the sum is always negative.
It is also possible to see from \afour\ that
\eqn\asix{ \eqalign {
\l 0 | T | \psi \r = {}& \sum_{n=0} a_n \big ( n - \sinh^2 \phi \big )
\l 0 | U |n\r \cr
= {}& (\cosh \phi)^{-{1\over 2}} \sum_{n=0} a_{2n}\, {\sqrt{(2n)!} \over 2^n
n!}
\,  \big ( 2n - \sinh^2 \phi \big ) \tanh^n\phi  \, . \cr}
}
Imposing the condition \ifive\ does not constrain $a_n$ for odd $n$ so that it
is
necessary to impose the condition
\eqn\aseven{
\sinh^2 \phi \le 1 \, ,
}
if the positivity condition \itwo\ subject to \ifive\ is to be at all
satisfied.
The condition \ifive\ may then be regarded as an equation determining $a_0$
giving
\eqn\aeight{
a_0 \,\sinh^2\phi  = \sum_{n=1} a_{2n}\,{\sqrt{(2n)!} \over 2^n n!}
\big ( 2n - \sinh^2 \phi \big ) \tanh^n\phi \, .
}
Using this to eliminate $a_0$ in \afive\ then after some algebra we find
\eqn\anine{ \eqalign{ \!\!\!\!
\l\psi |T|\psi \r {}& = \sum_{n=0} a_{2n+1}^{\,\, 2}
\big (2n+1 - \sinh^2 \phi \big ) \cr
&\  + \sum_{n=1} {2n+1\over 2n}\big (2n- \sinh^2\phi \big ) \tanh^2 \phi \cr
& \ \ \times \bigg (
a_{2n} - {1\over 2n+1}\,{2^n n!\over\sqrt{(2n)!}}\!
\sum_{m=n+1} \! a_{2m}\,{\sqrt{(2m)!}\over 2^m m!} \,
{2m- \sinh^2\phi \over \sinh^2\phi}
\tanh^{m-n}\phi \bigg )^{\! 2} \, , \cr}
}
which is obviously positive subject to \aseven.
\bigskip

An even simpler second illustration is given by considering the operator
\eqn\aten{
T = a^\dagger a + \lambda ( a+a^\dagger) = a'{}^\dagger a' - \lambda^2 \, ,
}
where
\eqn\aeleven{
a' = a + \lambda = U a U^{-1} \, , \quad U = e^{\lambda(a-a^\dagger)}
= e^{-{1\over 2}\lambda^2} e^{-\lambda a^\dagger} e^{\lambda a} \, .
}
As above a general state can be represented as $|\psi\r = \sum_{n=0} a_n U|n\r$
and then
\eqn\atwelve{
\l \psi | T | \psi \r = \sum_{n=0} a_n{}^{\! 2} ( n-\lambda^2) \, .
}
In this case imposing the restriction $\l 0 | T | \psi \r =0$ gives
\eqn\athirteen{
a_0 \lambda^2 = \sum_{n=1} a_n {1\over \sqrt{n!}}\,(n-\lambda^2)\lambda^n \, ,
}
and thus eliminating $a_0$ from \atwelve\ gives
\eqn\afourteen{
\l \psi | T | \psi \r = \sum_{n=1} { n-\lambda^2 \over n}\, \lambda^2
\bigg ( a_n  - \sqrt{n!}\! \sum_{m=n+1} a_m {1\over  \sqrt{m!}}\,
(m-\lambda^2) \, \lambda^{m-n-2} \bigg )^{\! 2} \, .
}
In consequence we have the positivity condition \itwo, subject to
\ifive, if
$\lambda^2<1$.

These examples demonstrate that the positivity conditions suggested here
may be valid for an operator $T$ which is close to an operator $T_0$
which annihilates $|0\r$. As the difference between $T$ and $T_0$ becomes
larger more and more conditions on the state $|\psi\r$ are necessary if
$\l \psi | T | \psi \r \ge 0$ is to be maintained.
\bigskip

\appendix{B}{}

In this appendix we show that conformal invariance ensures that there is no
loss
of generality in deriving positivity conditions for the states $|\psi_V\r$
and $|\psi_T\r$, defined in \inineteen\ and \ithirtysix,
 by setting $\bx={\bf 0}$. This
is essentially because a conformal transformation on Euclidean space allows
any three points to be made collinear so that we are able then to use
the simple expressions for the three-point functions given by
\itwentyfour\ and \ifourty.

To demonstrate this we first note that under a conformal transformation on
Euclidean space for which $x\to x'$, where
$\d x'{}_{\!\alpha}\d x'{}_{\!\alpha} = \Omega(x)^{-2} \d x_\alpha\d x_\alpha$,
a Euclidean vector field transforms as $V\to V'$ where
$V'{}_{\!\alpha}(x') = \Omega(x)^{d-1} \R(x)_{\alpha\beta} V_{\beta} (x)$
and we define the orthogonal matrix $\R$ by $\pr x'{}_{\!\alpha} / \pr x_\beta
= \Omega(x)^{-1}\R_{\alpha\beta}(x)$. For a special conformal transformation
\eqn\bone{
x'{}_{\!\alpha} = {x_\alpha + b_\alpha x^2 \over \Omega(x)} \, , \quad
\Omega(x)= 1 + 2b{\cdot x} + b^2 x^2 \, .
}
If $b_\alpha=(0,\bb)$ then the plane $x_\alpha \he_\alpha =0$ is left invariant
and we may choose $\bb = \bx/(\tau^2 + \bx^2)$ so as to transform
$(\hx,{\bf 0})\to (\tau,\bx)$. Assuming conformal invariance then from
 \itwentyfour\
we may obtain
\eqn\btwo{
\l  V^E{}_{\!\gamma}(\tau,\bx) \, V^E{}_{\!\de}(-\tau,\bx) \,
T^E{}_{\!\alpha\beta}(0)\r
= \R_{\gamma\gamma'}(\tau,\bx) \R_{\de\de'}(-\tau,\bx)\, {1 \over
(\tau^2 + \bx^2)^d (2\tau)^{d-2}} \, \A_{\gamma'\de'\alpha\beta} \, ,
}
where
\eqn\bthree{
\R_{\alpha\beta}(\tau,\bx)
 = \pmatrix { {\tau^2 - \bx^2 \over \tau^2 + \bx^2} &
- { 2\tau x_j \over \tau^2 + \bx^2} \cr {2\tau x_i \over \tau^2 + \bx^2} &
\de_{ij}  - {2 x_i x_j \over \tau^2 + \bx^2}} \ .
}
By taking into account the action of the matrix $\theta$ defined by
\isixteen\ we now find instead of \itwentyeight
\eqn\bfour{
\l \psi_V | n^\mu n^\nu T_{\mu\nu}(0) |\psi_V \r = {1\over (\tau^2+\bx^2)^d
(2\tau)^{d-2}} \, \psi^\si \psi^\rho \,{\tilde \R}_\si{}^{\si'}(\tau,\bx)^*
{\tilde \R}_\rho{}^{\rho'}(\tau,\bx) M_{\si'\rho'} \, ,
}
with now the matrix ${\tilde \R}$ given by
\eqn\bfive{
{\tilde \R}_{\mu}{}^{\nu}(\tau,\bx)
 = \pmatrix { {\tau^2 - \bx^2 \over \tau^2 + \bx^2} &
{2i\tau x^j \over \tau^2 + \bx^2} \cr {2i\tau x_i \over \tau^2 + \bx^2} &
\de_{i}{}^{\! j}  - {2 x_i x^j \over \tau^2 + \bx^2}} \ ,
}
which satisfies ${\tilde \R}_{\mu}{}^{\si}(\tau,\bx)
{\tilde \R}_{\mu}{}^{\rho}(\tau,\bx) g_{\si\rho} = g_{\mu\nu}$.
It is evident that positivity of \bfive\ also requires $M_{\si\rho}$ to be a
positive matrix just as \itwentyeight. It is also of interest to consider the
integral
over $\bx$ which gives
\eqn\bsix{ \eqalign{
\int \!  \d^{d-1}x \, & \l \psi_V | n^\mu n^\nu T_{\mu\nu}(0) |\psi_V \r  \cr
& =
2S_d {1\over (2\tau)^{2d-1}} {1\over d} \Big\{ (\psi^0)^2\big( M_{00}+M_{ii}
\big ) + 2\psi^0\psi^i \big( M_{0i} + M_{i0} \big ) + \psi^i \psi^i M_{00} \cr
& \qquad \qquad \qquad \qquad
+ {1\over d+1} \big ( \psi^i \psi^i M_{jj} + d(d-1) \psi^i \psi^j M_{ij}
\big ) \Big \} \, . \cr}
}
If we set $\bn ={\bf 0}$, so that $M_{0i} = M_{i0}=0, \ M_{ij} \propto
\de_{ij}$, then it is easy to see that from \itwentytwo
\eqn\bseven{
\l \psi_V | H |\psi_V \r = 2S_d {1\over (2\tau)^{2d-1}} \, {1\over d} \,
\psi^\mu\psi^\mu M_{\si\si}= -\half {\pr \over \pr \tau}\l \psi_V|\psi_V \r \,
,
}
if, using \itwentytwo,
\eqn\beight{
(d-1)C_V = {S_d\over d}\, M_{\si\si} \, .
}
This is of course in agreement with the Ward identity result \itwentyseven.
It is perhaps
worth noting that the unitarity condition $C_V >0$ constrains the trace of
the matrix $M_{\si\rho}$ while positivity of the matrix element of $T_{00}$
gives positivity of the whole matrix when $\bn ={\bf 0}$. An additional
constraint on the matrix may be obtained by considering the momentum operator
\eqn\bnine{
P_i = - \int \!  \d^{d-1}x \, T_{0i}(0,\bx) \, .
}
By expanding \bsix\ to first order in $\bn$, and using \itwentynine, we may
find
\eqn\bten{
\l \psi_V | {\bf P}|\psi_V \r = 2S_d {1\over (2\tau)^{2d-1}} \, {4\over d}\,
\delta \, \psi^0 {\bold \psi} = -4C_V {1\over (2\tau)^{2d-1}}\,
\psi^0 {\bold \psi} \, ,
}
where the second expression results by using ${\bf P}$ as a generator of
translations in $\bx$. Clearly \bten\ requires $2S_d \delta = -d C_V$ in
agreement with \itwentysix,\itwentyseven.

In a similar fashion for the state $|\psi_T \r$ defined in \ithirtysix\ for
arbitrary $\bx$ we find
\eqn\beleven{ \eqalign { \!\!\!\!\!\!\!\!\!\!\!
\l & \psi_T |  n^\mu n^\nu T_{\mu\nu}(0) |\psi_T \r \cr \!\!\!\!
& = {1\over (\tau^2+\bx^2)^d(2\tau)^d }\, \psi^{\si\rho}\psi^{\kappa\lambda}\,
{\tilde \R}_\si{}^{\si'}(\tau,\bx)^* {\tilde \R}_\rho{}^{\rho'}(\tau,\bx)^*
{\tilde \R}_\kappa{}^{\kappa'}(\tau,\bx)
{\tilde \R}_\lambda{}^{\lambda'}(\tau,\bx)
M_{\si'\rho',\kappa'\lambda'} \, , \cr}
}
and furthermore, instead of \ithirtynine, we now have
\eqn\btwelve{
\l 0 | n^\mu n^\nu T_{\mu\nu}(0) |\psi_T \r = C_T \,
{1\over (\tau^2 + \bx^2)^d}\, n^\mu n^\nu
{\tilde \R}_\mu{}^{\si}(\tau,\bx){\tilde \R}_\nu{}^{\rho}(\tau,\bx)
\psi^{\si\rho} \, .
}
Clearly we gain no more from positivity conditions for $\bx \ne {\bf 0}$
than we have found from \ifourtyseven. From \btwelve\ we then have
\eqn\bthirteen{
\int \!  \d^{d-1}x \, \l 0 | n^\mu n^\nu T_{\mu\nu}(0) |\psi_T \r
= C_T \, {2S_d \over d+1} \big ( (d-1) n^i n^j \psi^{ij} - \bn^2 \psi^{ii}
\big ) \, ,
}
which is in accord with the requirement that $\l 0 | P^\mu |\psi_T \r =0$
where $P^\mu = (H,{\bf P})$.
\bigskip

\bigskip\bigskip
\listrefs
\bye